# Giant modification of atomic transitions probabilities induced by magnetic field: forbidden transitions become predominant

A. Sargsyan[1], A. Tonoyan[1], G. Hakhumyan[1], A. Papoyan[1], E. Mariotti[2], D. Sarkisyan[1]

[1] Institute for Physical Research, NAS of Armenia, Ashtarak-2, 0203, Armenia
[2] CNISM – Department of Physical, Earth and Environmental Sciences, University of Siena, Via Roma 56, 53100 Siena, Italy

**ABSTRACT**

Magnetic field-induced giant modification of probabilities for seven components of $6S_{1/2}$ ($F_g=3$) → $6P_{3/2}$ ($F_e=5$) transition of Cs $D_2$ line forbidden by selection rules is observed experimentally for the first time. For the case of excitation with circularly-polarized laser radiation, the probability of $F_g=3, m_F=-3$ → $F_e=5, m_F=-2$ transition becomes the largest among 25 transitions of $F_g=3$ → $F_e=2,3,4,5$ group in a wide range of magnetic field 200 - 3200 G. Moreover, the modification is the largest among $D_2$ lines of alkali metals. A half-wave-thick cell (length along the beam propagation axis L=426 nm) filled with Cs has been used in order to achieve sub-Doppler resolution which allows for separating the large number of atomic transitions that appear in the absorption spectrum when an external magnetic field is applied. For B > 3 kG the group of seven transitions Fg=3 → Fe=5 is completely resolved and is located at the high frequency wing of Fg=3 → Fe=2,3,4 transitions. The applied theoretical model very well describes the experimental curves.

**INTRODUCTION**

Alkali atoms are widely used in atomic physics due to simplicity of electronic structure and strong atomic transitions from the ground state with wavelengths in visible and near-infrared, where diode lasers with good parameters are available. Cesium atoms are widely used in laser cooling experiments, information storage, spectroscopy, magnetometry, laser frequency stabilization etc. [1-3]. That is why any knowledge of the behavior of Cs atomic transitions, particularly, in an external magnetic field is of high importance. It is well known that in quite moderate magnetic field *B* the splitting of atomic energy levels to Zeeman sublevels deviates from the linear behavior, and the atomic transition probabilities undergo significant changes [4-6]. The most simple and straightforward technique to study such modification of atomic transitions (whose frequency distance belongs to optical domain) is laser spectroscopy of atoms contained in an atomic vapor cell. For *B* up to $\cong$ 1000 G, the split Zeeman transitions remain overlapped because of Doppler broadening, and sub-Doppler techniques have to be implemented in order to spectrally resolve and study transition probability of individual transition components [7].

As it was demonstrated earlier, strong narrowing in absorption spectrum can be attained with the use of an atomic vapor cell of half-wavelength thickness ($L = \lambda/2$, where $\lambda$ is the resonant wavelength of laser radiation, $L = 426$ nm for the case of Cs $D_2$ line) [8-11]. Particularly, the absorption linewidth for Cs $D_2$ line reduces to $\cong$ 100 MHz (FWHM), as opposed to $\cong$ 400 MHz in an ordinary cell. Moreover, the absorption lines for $L = \lambda/2$ exhibit Voigt profile (a convolution of Lorentzian and Gaussian profiles) with sharp (nearly Gaussian) peak, which allows separation of closely spaced individual transitions and study their transition probabilities in an external magnetic field. In addition, the $\lambda/2$-method is tolerant against 10% deviation of thickness (weak influence on the absorption linewidth). These benefits make it convenient to use $\lambda/2$-method for studies of closely spaced individual atomic transition components in a magnetic field.

In this letter we present, for the first time, the results of experimental and theoretical studies showing a giant transition probability modification for Cs $D_2$ line $6S_{1/2}$, $F_g=3 \rightarrow 6P_{3/2}$, $F_e=5$ transition induced by a magnetic field varied in a wide range up to 3.5 kG. It should be remembered that, according to the electric dipole selection rules in zero B field, only transitions with a change of atomic total angular momentum $\Delta F = F_g - F_e = 0, \pm 1$ are allowed, while $F_g=3 \rightarrow F_e=5$ transitions are forbidden. To the best of our knowledge, there are only few articles where such type of transitions have been quantitatively studied [4,6,11].

**EXPERIMENTAL**

Nanometric Thin Cells (NTCs) filled with Cs have been used in our experiment, because of their peculiar properties which allow a) to directly obtain sub-Doppler spectra without application of nonlinear techniques and therefore b) to resolve even complicated spectra, as already demonstrated in many papers published in the last few years after their introduction in laser spectroscopy [8].

The general design of NTC is similar to that described in ref.12. Compact oven has been used to set the needed temperature regime. The temperature was set to 100 $^o$C, which corresponds to number density of isotopically pure $^{133}$Cs atoms $N \cong 10^{13}$ cm$^{-3}$. Adjustment of needed vapor column thickness without variation of thermal conditions was attained by smooth vertical translation of the cell+oven assembly.

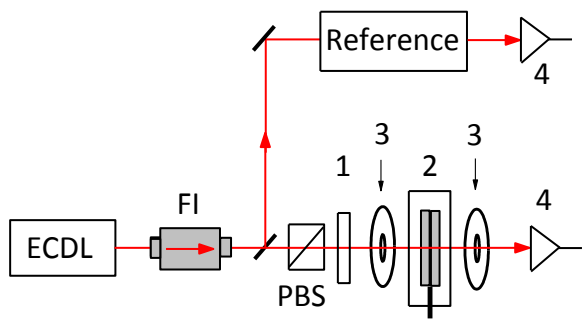

Fig.1. Sketch of the experimental setup. ECDL- External Cavity Diode Laser; FI- Faraday Isolator; 1- $\lambda/4$ plate; 2- NTC in the oven; PBS- Polarizing Beam Splitter; 3- permanent ring magnets; 4- photodetectors; Reference- auxiliary Cs NTC providing $B = 0$ reference spectrum.

Schematic diagram of the optical part of experimental setup is shown in Fig.1. A circularly polarized laser radiation beam ($\lambda$ = 852 nm, $P_L$ = 5 mW, $\Delta\nu_L$ = 1 MHz) resonant with Cs $D_2$ line was focused ($\varnothing$ = 0.5 mm) onto a Cs NTC with a vapor column of thickness $L = \lambda/2$ at normal incidence angle. The 8 mm-thick assembly of the oven with main NTC inside was placed between two permanent ring magnets (whose axis is directed along the laser radiation propagation direction **k**) with gradually adjustable spacing providing controllable longitudinal $B$-field (for details, see [6]). Extremely small thickness of NTC is once more advantageous for application of very strong magnetic field with the use of permanent magnets otherwise unusable because of strong inhomogeneity: in NTC, the variation of the $B$-field inside the cell is several orders less than the applied $B$ value [13]. To record transmission and fluorescence spectra, the laser radiation was linearly scanned within up to 15 GHz spectral region covering the studied group of transitions. The nonlinearity of the scanned frequency (< 1 % throughout the spectral range) was monitored by simultaneously recorded transmission spectra of a Fabry–Pérot etalon (not shown). About 30% of the pump power was branched to an auxiliary Cs NTC with thickness $L = \lambda/2$ providing reference absorption spectrum for $B = 0$. All the spectra were detected by photodiodes with amplifiers followed by a four channel digital storage oscilloscope Tektronix TDS 2014B.

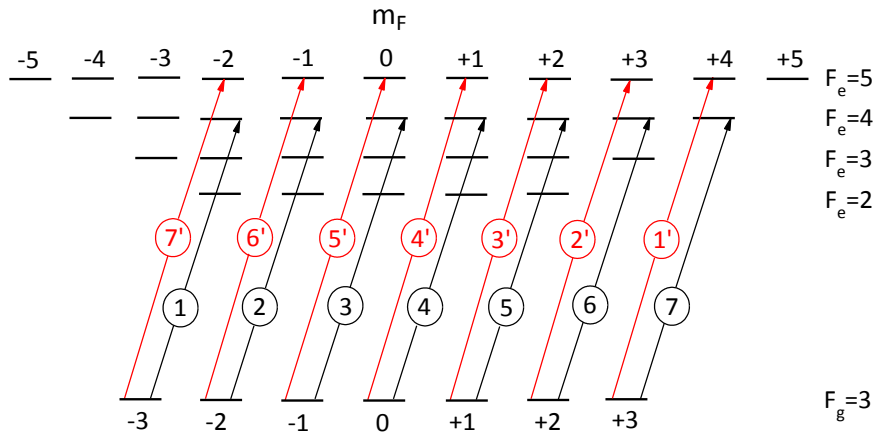

Fig.2. Diagram of relevant transitions between Zeeman sub-levels of ground ($F_g=3$) and excited ($F_e=4,5$) states for $D_2$ line of $^{133}$Cs (nuclear spin I = 7/2) with σ+ (left-hand) laser excitation. Labeling of transitions is given in circles.

The diagram of relevant σ$^+$ components ($\Delta m_F = +1$) of Cs $D_2$ line transitions $F_g=3 \rightarrow F_e=4$ labeled 1 - 7, and $F_g=3 \rightarrow F_e=5$ labeled 1′ - 7′ is shown in Fig.2. The transitions 1′ - 7′ are forbidden for zero magnetic field because of selection rule $\Delta F = 0,\pm 1$. The transitions $F_g=3 \rightarrow F_e=2,3$ are not shown, since for $B > 500$ G their probabilities strongly reduce, and these transitions are practically not detectable in the absorption spectrum.

The recorded absorption spectrum of Cs NTC with thickness $L = \lambda/2$ for σ$^+$ laser excitation ($P_L = 10$ μW) and $B = 920$ G is shown in Fig.3. The fourteen above labeled transition components appear with ≅ 100 MHz linewidth, thus being completely frequency resolved except for transitions 5, 6 and 7′ resolved partially, and 6′ and 7, which are fully overlapped. As it is seen from the inset the amplitudes of Fg=3, m$_F$=-3→ Fe=5, m$_F$=-2 and Fg=3, m$_F$=-2→ Fe=5, m$_F$=-1 transitions (labeled 7′ and 6′, respectively) are the largest ones among the group of fourteen lines.

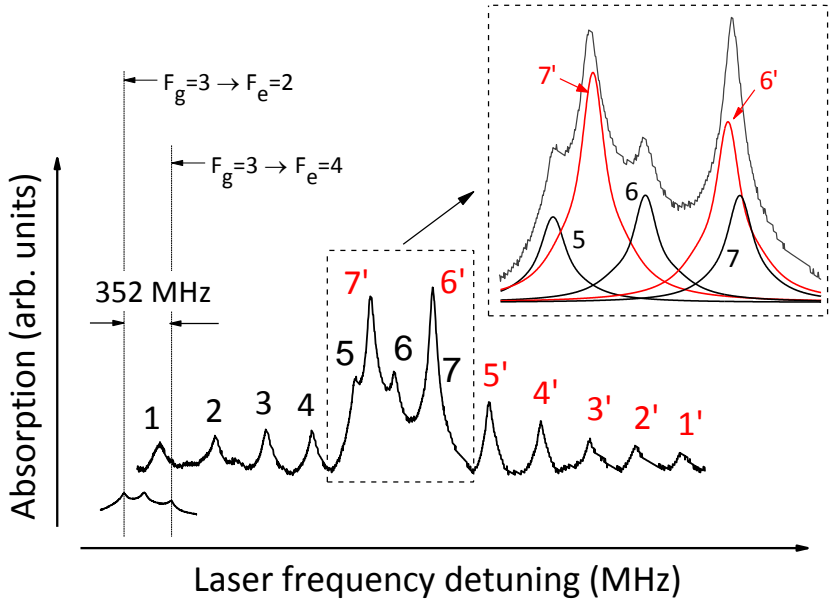

Fig.3. Absorption spectrum of Cs NTC with $L = \lambda/2$ for $B = 920$ G and σ$^+$ laser excitation. For labeling of transitions, see Fig.2. The bottom-left curve is absorption spectrum of the reference NTC showing positions of $F_g=3 \rightarrow F_e=2,3,4$ transitions for $B = 0$. The inset shows an expanded view of the part of the experimental results limited by the dashed rectangle and the corresponding fit of the individual lines.

The fitting is justified thanks to the following advantageous property of the λ/2-method: for the case of a weak absorption, the absorption coefficient of an individual transition component $A$ is proportional to σ$NL$, where σ is the absorption cross-section proportional to $d^2$ ($d$ being the matrix element of the dipole moment), $N$ is atomic density, and $L$ is the thickness. Measuring the ratio of $A_i$ values for different individual transitions, it is straightforward to estimate their relative probabilities (line intensities). As is seen from the inset, the amplitudes of 7′ and 6′ transitions forbidden at $B = 0$ are the largest among all fourteen transitions.

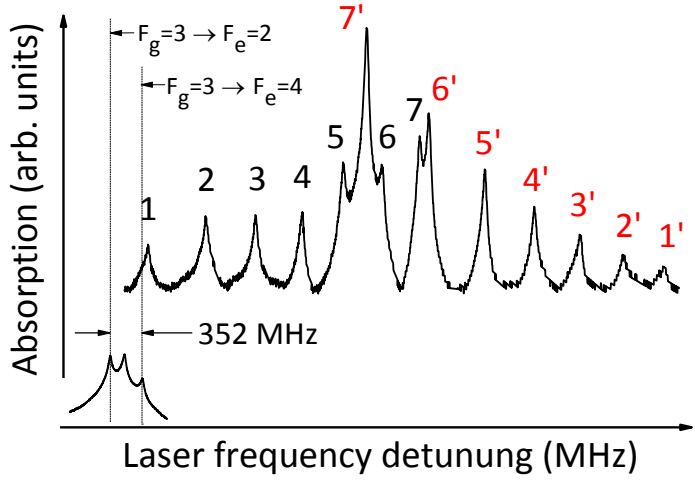

Fig.4. Absorption spectrum of Cs NTC with $L = \lambda/2$ for $B = 1450$ G and $\sigma^+$ laser excitation. For labeling of transitions, see Fig.2. The bottom-left curve is absorption spectrum of the reference NTC showing positions of $F_g=3 \rightarrow F_e=2,3,4$ transitions for $B = 0$.

Further increase of magnetic field to $B = 1450$ G results in better resolution of individual transitions (Fig.4). Complete separation of all the fourteen transition components occurs at $B > 3000$ G. The absorption spectrum for $B = 3450$ G and otherwise invariable experimental conditions as in Fig.3 and Fig.4 is presented in Fig.5. Here also the groups of seven transitions 1 - 7 ($F_g=3 \rightarrow F_e=4$) and 1′ - 7′ ($F_g=3 \rightarrow F_e=5$) are completely separated as marked by dashed rectangles. In addition, two transition components from the group $F_g=4 \rightarrow F_e=5$ appear to be located on the low frequency side of the spectrum.

Noteworthy that the intensity of the transition component labeled 7′ is the largest among all the 25 atomic transitions of $F_g=3 \rightarrow F_e=2,3,4,5$ group for the magnetic field range 250 G $< B <$ 3200 G.

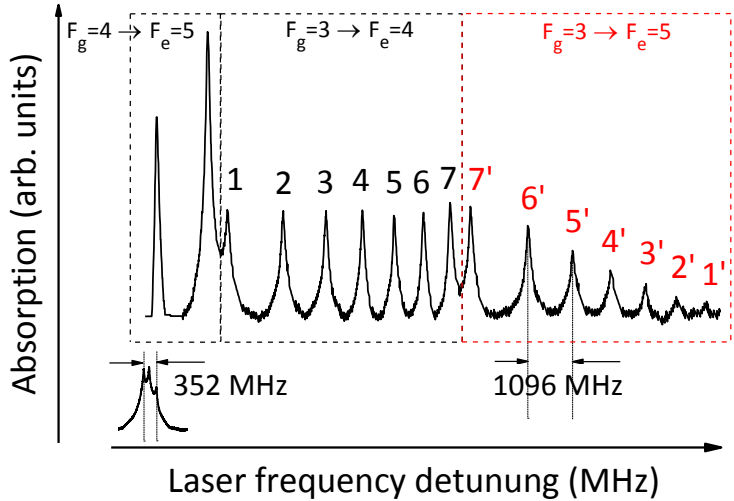

Fig.5. Absorption spectrum of Cs NTC with $L = \lambda/2$ for $B = 3450$ G and $\sigma^+$ laser excitation. For labeling of transitions, see Fig.2. The bottom-left curve is absorption spectrum of the reference NTC showing positions of $F_g=3 \rightarrow F_e=2,3,4$ transitions for $B = 0$. All the individual Zeeman transitions 1–7 and 1′–7′ are completely resolved. Two shifted components of $F_g=4 \rightarrow F_e=5$ transition appear at the low frequency region.

**THEORETICAL MODEL AND DISCUSSION**

Simulations of magnetic sublevel energy and relative transition probabilities for $F_g=3 \rightarrow F_e=2,3,4,5$ transitions of Cs $D_2$ line are well known, and are based on the calculation of dependence of the eigenvalue and eigenvector of the Hamilton matrix on magnetic field for the full hyperfine structure manifold [2,4,5]. The dependence of frequency shifts of transitions 1′ - 7′ and 1 - 7 on magnetic field relative to position of $F_g=3 \rightarrow F_e=4$ transition at $B = 0$ for the case of $\sigma^+$ excitation is shown in Fig.6. Good agreement of theory and experiment is observed throughout the whole explored range of $B$-field (up tp 3500 G).

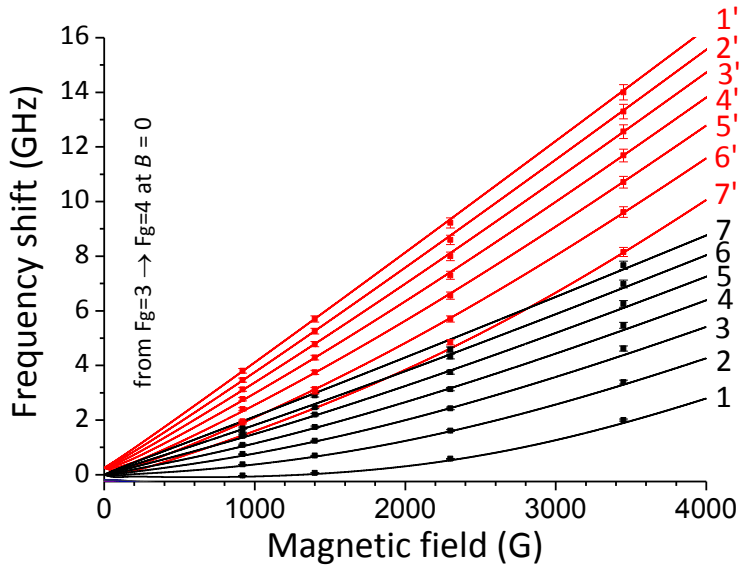

Fig.6. Magnetic field dependence of frequency shift for atomic transition labeled 1′ - 7′ and 1 - 7 relative to the position of $F_g=3 \rightarrow F_e=4$ transition at $B = 0$. Black squares: experimental results (inaccuracy $\cong 2\ \%$); solid curves: calculated dependence.

The calculated dependence of 1′ - 7′ and 1 - 7 transition probabilities (absorption amplitudes) on magnetic field for the case of $\sigma^+$ laser excitation are shown in Fig.7.

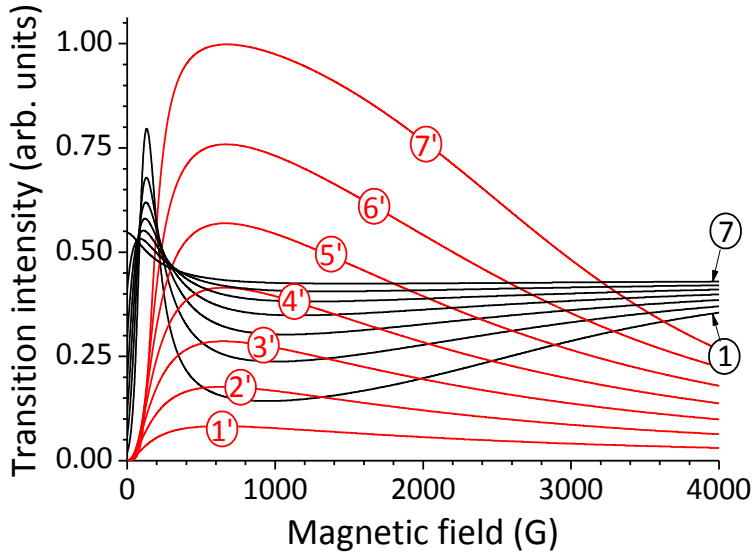

Fig.7. Calculated probabilities (absorption amplitudes) of transitions 1′ - 7′ and 1 - 7 versus magnetic field for $\sigma^+$ laser excitation.

Since the absolute value of the absorption amplitude $A$ depends on parameters of the experiment (laser intensity, atomic density, etc.), it is expedient to present also the $B$-field dependence of the ratio of absorption amplitudes $A_i^{'}$ of 1′ - 7′ transitions to absorption amplitude $A_7$ of the transition 7. The latter is the strongest in the $F_g=3 \rightarrow F_e=4$ group; moreover, the absorption amplitude $A_7$ is nearly constant in a wide range 250 G < $B$ < 4000 G (see Fig.7), which makes it convenient to use as a reference. The theoretical ratio $A_i^{'} / A_7$ versus $B$-field is plotted in Fig.8 together with experimental results (the ratio is easily measurable). The dashed line marks the unity ratio. As it is seen from Fig.8, $A_7' / A_7 > 1$ holds in a wide range 200 G < $B$ < 3200 G, and the maximum value of the ratio is 2.3.

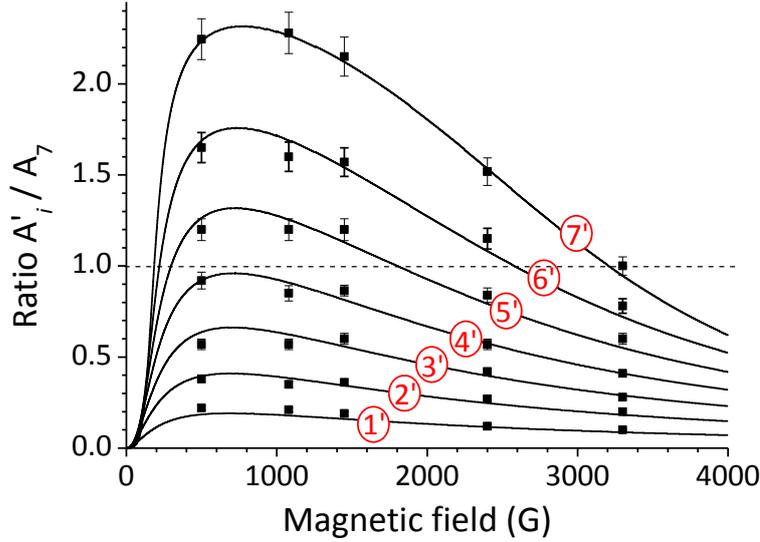

Fig.8. Ratio of absorption amplitudes $A_i^{'}$ (transitions 1′ - 7′) to $A_7$ versus $B$ for the case of $\sigma^+$ excitation. The dashed line shows the range of magnetic field where the ratio $A_i^{'} / A_7 =1$.

Thus, the transition 7′, as well as 6′ and 5′, which are forbidden for $B = 0$ undergo giant modification under the influence of magnetic field, becoming predominant over the initially allowed transitions 1 - 7. It is worth noting that the maximum value of transition probability for 7′ transition reaches 76 % of the probability for $F_g=4, m_F=+4 \rightarrow F_e=5, m_F=+5$ transition, which is the strongest among all the 54 Zeeman transitions of Cs $D_2$ line.

We want also to stress that another group of transitions forbidden at $B = 0$, $F_g=4 \rightarrow F_e=2$ (5 transition components for the case of $\sigma^+$ excitation), also exhibits modification induced by magnetic field. However, its maximum probability value achieved at $B = 60$ G is over 30 times smaller than the maximum probability value for $F_g=3 \rightarrow F_e=5$.

The decoupling of the total angular momentum $J$ and the nuclear momentum $I$ (hyperfine Paschen-Back regime) for Cs atom occurs for $B >> B_0 = A_{HFS}/\mu_B \cong 1700$ G, where $A_{HFS} = h \times 2.3$ GHz is the magnetic dipole hyperfine constant for $6S_{1/2}$, and $\mu_B$ is the Bohr magneton. In this case the splitting of transitions is described by the projections $m_J$ and $m_I$ [13–18]. At $B > 6000$ G, sixteen transitions are observable in the absorption spectrum: by 8 starting from the ground states $6S_{1/2}, m_J=-1/2$ and and $6S_{1/2}, m_J=+1/2$. For $B > 3000$ G and $\sigma^+$ excitation, the group of $F_g=3 \rightarrow F_e=5$ transitions it is always located at the high frequency side of $F_g=3 \rightarrow F_e=2,3,4$ transitions, with line intensities monotonically reducing and completely vanishing at $B > 9000$ G.

It is interesting to compare the evolution of maximum probability value of magnetic field-induced Cs $F_g=3 \rightarrow F_e=5$ transition normalized to the strongest $D_2$ transition with the corresponding values for $D_2$ line transitions forbidden at $B = 0$ for other alkali metal atoms. The theoretical calculations show that for Rb transitions $^{85}$Rb $5S_{1/2}, F_g=2 \rightarrow 5P_{3/2}, F_e=4$ and $^{87}$Rb $5S_{1/2}, F_g=1 \rightarrow 5P_{3/2}, F_e=3$, the maximum value is less 1.10 and 1.36 times, correspondingly. Also for $3S_{1/2}, F_g=1 \rightarrow 3P_{3/2}, F_e=3$ transitions of Na and $4S_{1/2}, F_g=1 \rightarrow 4P_{3/2}, F_e=3$ of K the maximum values are less (1.31 and 1.33 times, correspondingly). Thus, the modification of the probabilities for cesium $F_g=3 \rightarrow F_e=5$ transitions is the strongest among all the alkali atoms.

## CONCLUSION AND OUTLOOK

Giant modification of probabilities of Cs $D_2$ line transitions $6S_{1/2}, F_g=3 \rightarrow 6P_{3/2}, F_e=5$ induced by magnetic field $B$ has been studied both the experimentally and theoretically. It has been revealed for the first time that the absorption intensity for $F_g=3, m_F=-3 \rightarrow F_e=5, m_F=-2$ transition becomes the largest among all the 25 Zeeman transitions of the $F_g=3 \rightarrow F_e=2,3,4,5$ manifold for the case of $\sigma^+$ excitation in a wide range of magnetic field 200 - 3200 G. It is demonstrated that utilization of a half-wavelength-thick cell filled with Cs, favorable for strong reduction of Doppler broadening of absorption lines, allows to quantitatively study the frequency positions and modification of individual transition probabilities. As a result of special interest, for $B > 3000$ G all the seven transitions of $F_g=3 \rightarrow F_e=5$ group are completely separated. Calculated theoretical curves for transition frequency shifts and modification of probabilities induced by magnetic field for all the transitions under study show very good coincidence with the experimental results.

The largest probabilities of "forbidden" transitions labeled 7′, 6′ and 5′ in a wide range of magnetic field (see Fig.8) make them attractive for formation of sub-natural Electromagnetically Induced Transparency resonances in new frequency regions [3,19] (for the realization of this experiment, the coupling laser frequency must be in resonance with $F_g=4 \rightarrow F_e=5$, which is easy to realize). It should be noted that λ/2-M method can be implemented successfully to study forbidden transitions of $D_2$ lines of other alkalis, as Rb, K and Na.

**Acknowledgements**

The authors are grateful to A. Sarkisyan for his valuable participation in the development and fabrication of the NTC. This work has received partial funding from the EU Seventh Framework Programme (FP7/2007-2013) under Grant Agreement n° 295264-COSMA. A.S., G.H. and D.S. acknowledge financial support from the State Committee Science, MES of Armenia (Research Project № 13-1C029).